\documentclass{siamltex}
\usepackage{color}
\usepackage{graphicx} 
\usepackage{here}     
\usepackage{amsmath}
\usepackage{psfrag}
\usepackage{dsfont}

\title{On the efficiency and accuracy of interpolation methods for spectral codes}

\author{M.A.T. van Hinsberg\footnotemark[2]
        \and J.H.M. ten Thije Boonkkamp\footnotemark[3]
         \and F. Toschi\footnotemark[2]\ \footnotemark[3]\ \footnotemark[4]
         \and H.J.H. Clercx\footnotemark[2]\ \footnotemark[5]}

\begin{document}

\maketitle

\renewcommand{\thefootnote}{\fnsymbol{footnote}}
\footnotetext[2]{Department of Physics,
    Eindhoven University of Technology, PO Box 513,
    5600MB Eindhoven, The Netherlands}
\footnotetext[3]{Department of Mathematics and Computer Science,
     Eindhoven University of Technology, PO Box 513,
     5600MB Eindhoven, The Netherlands}
\footnotetext[4]{CNR, Istituto per le Applicazioni del Calcolo, Via dei Taurini 19,
00185 Rome, Italy}
\footnotetext[5]{Department of Applied Mathematics,
    University of Twente, PO Box 217, 7500 AE Enschede, The Netherlands}
\renewcommand{\thefootnote}{\arabic{footnote}}

\begin{abstract}
In this paper a general theory for interpolation methods on a rectangular grid is introduced. By the use of this theory an efficient B-spline based interpolation method for spectral codes is presented. The theory links the order of the interpolation method with its spectral properties. In this way many properties like order of continuity, order of convergence and magnitude of errors can be explained. Furthermore, a fast implementation of the interpolation methods is given. We show that the B-spline based interpolation method has several advantages compared to other methods. First, the order of continuity of the interpolated field is higher than for other methods. Second, only one FFT is needed whereas e.g. Hermite interpolation needs multiple FFTs for computing the derivatives. Third, the interpolation error almost matches the one of Hermite interpolation, a property not reached by other methods investigated.
\end{abstract}



\pagestyle{myheadings}
\thispagestyle{plain}

\section{Introduction}

In recent years many studies on the dynamics of inertial particles in turbulence have focussed on the Lagrangian properties, see the review by Toschi and Bodenschatz \cite{Toschi3}. For particles in turbulence, but also in many other applications in fluid mechanics, interpolation methods play a crucial role as fluid velocities, rate of strain and other flow quantities are generally not available at the location of the particles, while these quantities are needed for the integration of the equations of motion of the particles.

When a particle is small, spherical and rigid its dynamics in non-uniform flow is governed by the Maxey-Riley (MR) equation \cite{Maxey}. An elaborate overview of the different terms in the MR equation and their numerical implementation can be found in the paper by Loth \cite{loth00} and a historical account was given in a review article by Michaelides \cite{Michaelides2}. The evaluation of the hydrodynamic force exerted on the particles requires knowledge of the fluid velocity, its time derivative and gradients at the particle positions and turns out to be rather elaborate. First, the Basset history force is computationally very expensive. However, a significant reduction of cpu-time can be obtained by fitting the diffusive kernel of the Basset history force with exponential functions, as recently shown by Van Hinsberg \textsl{et al.} \cite{Hinsberg}. Second, the interpolation step itself can be very time consuming and memory demanding. Especially for light particles, which have a mass density similar to the fluid density (which is, for example, sediment transport in estuaries and phytoplankton in oceans and lakes), most terms in the Maxey-Riley equation cannot be ignored and therefore also the first derivatives of the fluid velocity are needed \cite{Marleen2}. For this reason simulations of light particles are computationally expensive while simulations of heavy particles are less expensive. In order to achieve convergence of the statistical properties (probability distribution functions, correlation functions, multi-particle statistics, particle distributions) many particles are needed and this calls for fast and accurate interpolation methods. Therefore, our aim is to reduce the computation time for the evaluation of the trajectories of light particles substantially and make the algorithm competitive with the fast algorithms for the computation of trajectories of heavy particles in turbulence.

The incompressible Navier-Stokes equations are used to describe the turbulent flow field. In turbulence studies the Navier-Stokes equations are often solved by means of dealiased pseudo-spectral codes because of the advantage of exponential convergence of the computed flow variables. Therefore, we will focus here on interpolation methods for spectral codes.

There are many interpolation methods available \cite{Medical1}. We are interested in those interpolation methods which are characterized by the following properties. First, the method must be accurate, thus we need a high order of convergence. Second, the interpolant must have a high order of continuity $C^p$, with $p$ the order of continuity. Third, the method must be fast, i.e. computationally inexpensive. A very simple interpolation method is linear interpolation. This method is very fast, but unfortunately this method is relatively inaccurate and it has a low order of convergence. High order of convergence can be reached by employing Lagrange interpolation \cite{Faires}. This interpolation method has the drawback that it still has a low order of continuity for the interpolant. A solution for this problem was recently found by Lalescu \textsl{et al.} \cite{Lalescu} who proposed a new spline interpolation method. Here, the interpolant has a higher order of continuity, but the order of convergence has decreased. A method that has both a high order of convergence and a high order of continuity is Hermite interpolation \cite{Hermite}. The major disadvantage of this method is that also the derivatives of the function to be interpolated are needed, these are calculated by additional Fast Fourier Transforms (FFTs) making this method computationally expensive. A remedy to this is B-spline interpolarion \cite{Bspline2}, which has a high order of convergence and errors comparable with the ones of Hermite interpolation. Furthermore, this method has a higher order of continuity compared with the other methods mentioned above. Normally, the transformation to the B-spline basis is an expensive step, but by making use of the spectral code it can be executed in Fourier space which makes it inexpensive. By executing this step in Fourier space the method can be optimized, resulting in smaller errors. We believe that the proposed combination of B-spline interpolation with a spectral code makes the method favorable over other traditional interpolation methods.

Besides exploring the advantages of B-spline interpolation we focus on necessary conditions allowing general 3D-interpolations to be efficiently executed as successive 1D-interpolations.  These conditions also carry over desired properties (like order of convergence and order of continuity) from the 1D-interpolation to the three-dimensional equivalent. Further, we provide a fast, generic algorithm to interpolate the function and its derivatives using successive 1D-interpolations.

We provide expressions for the interpolation errors in terms of the Fourier components. For this we use Fourier analysis where the interpolation method is represented as a convolution function. By doing this, the errors can be calculated as a function of the wave number. This gives insight in the behavior of interpolation, especially which components are dominant in the interpolation.

The present study may also be useful for many other applications. Examples include the computation of charged particles in a magnetic field \cite{Reeves,Mackay}, but also digital filtering and applications in medical imaging \cite{Medical1,medical2}. In the latter case interpolations are used to improve the resolution of images. Many efforts have been taken to find interpolation methods with optimum qualities \cite{Medical1}. Still, it is a very active field of research. Besides the optimization of interpolation algorithms (accuracy, efficiency), the impact of different interpolation methods on physical phenomena like particle transport has been investigated in many studies \cite{Homann2,choi,marchioli}.

In Section \ref{sec_met} we introduce the general framework and explain some one-dimen-sional interpolation methods. In Section \ref{sec_3d} the framework is generalized to three-dimensional interpolation, and a generic algorithm is proposed for the implementation of the interpolation in Section \ref{sec_implementation}. A Fourier analysis of the interpolation operator is discussed in Section \ref{sec_fourier}. In Section \ref{sec_hermite} the Fourier analysis is extended to Hermite interpolation and a proof of minimal errors is given. In Section \ref{sec_B_spline} our B-spline based interpolation method is introduced, and is compared with three other methods (including Hermite interpolation) in Section \ref{sec_comparison}. Finally, concluding remarks are given in Section \ref{sec_conclusion}.
\section{Interpolation methods}\label{sec_met}
We present a general framework to discuss the various interpolation methods. The goal of any interpolation method is to reconstruct the original function as closely as possible. As in many applications also some derivatives of the original function are needed, we focus on reconstructing them as well. We start with one-dimensional (1D) interpolation and subsequently, in Section \ref{sec_3d}, we generalize our framework to the three-dimensional (3D) case.

Let $u(x)$ be a 1D function that needs to be reconstructed with $x\in[0,1]$. In practice we only have the values of $u$ on a uniform grid, with grid spacing $\Delta x$ and knots at positions $x_j$, with $0\leq j\leq(\Delta x)^{-1}$. After interpolation, the function $\widetilde{u}$ is obtained which is an approximation of $u$. Now let $\mathcal{I}$ be the interpolation operator, so $\widetilde{u}=\mathcal{I}[u]$.

When $u$ has periodic boundary conditions, it can be expressed in a Fourier series as follows
\begin{eqnarray}
u(x)=\sum_{k\in \mathds{Z}}\hat{u}_k \phi_k(x), ~~~~~~
\phi_k(x)=e^{2\pi \textrm{i} k x},\label{eq_phi_k}
\end{eqnarray}
with $\textrm{i}$ the imaginary unit and $k$ the wave number. As the grid spacing is finite, a finite number of Fourier modes can be represented by the grid. From now on we consider $u$ to have a finite number of Fourier modes, so
\begin{eqnarray}
u(x)=\sum_{k=-k_{\textrm{max}}}^{k_{\textrm{max}}}\hat{u}_k \phi_k(x),
\end{eqnarray}
where $k_{\textrm{max}}$, related to $\Delta x$, is the maximum wave number. As we add a finite number of continuously differentiable Fourier modes $\phi_k$ we have $u\in C^\infty(0,1)$, a property which can be used when constructing the interpolation method. In principle $u$ could be reconstructed at any point by the use of Fourier series, however in practice this would be far too time consuming and it is therefore not done, instead an interpolation is performed. $\widetilde{\phi}_k$ is defined as the interpolant of $\phi_k$, i.e.,
$\widetilde{\phi}_k=\mathcal{I}\left[\phi_k\right]$.
We restrict ourselves to linear interpolation operators, i.e., $\mathcal{I}\left[\alpha_1u_1+\alpha_2u_2\right]=\alpha_1\mathcal{I}\left[u_1\right]+\alpha_2\mathcal{I}\left[u_2\right]$ with $\alpha_1,\alpha_2\in\mathds{C}$. This property can be used to write $\widetilde{u}(x)$ as
\begin{eqnarray}
\widetilde{u}(x)=\sum_{k=-k_{\textrm{max}}}^{k_{\textrm{max}}}\hat{u}_k \widetilde{\phi}_k(x).
\end{eqnarray}

We focus on reconstructing $u$ by piecewise polynomial functions of degree $N-1$. For each interval $(x_j,x_{j+1})$ with $0\leq j<(\Delta x)^{-1}$ we have


\begin{eqnarray}
\widetilde{u}(x)=\sum_{i=1}^{N}a_ix^{i-1}=\textbf{a}^T\bar{\textbf{x}},~~~~~~x\in\left(x_j,x_{j+1} \right),~~~~~~\bar{\textbf{x}}=\left( \begin{array}{c}
   1 \\
   x \\
  x^2 \\
  \vdots\\
  x^{N-1} \\
  \end{array}\right).\label{eq-vector-a}
\end{eqnarray}
Here, the vector $\textbf{a}$ depends on the interval under consideration and $\textbf{a}^T$ denotes the transpose of $\textbf{a}$. The degree of the highest polynomial function for which the interpolation is still exact is denoted by $n$. In this way we get the restriction $n\leq N-1$. We consider the reconstruction of $u$ between the two neighboring grid points $x_j$ and $x_{j+1}$. Without loss of generality we can translate and rescale $x$ so that the interval $[x_j,x_{j+1}]$ becomes the unit interval $[0,1]$.

For Hermite interpolation the values of $\widetilde{u}$ and of its derivatives, up to the order $N/2-1$  ($N$ must be even), must coincide with those of $u$ at $x=0$ and $x=1$, i.e.,
\begin{eqnarray}
\frac{\textrm{d}^{l}\widetilde{u}}{\textrm{d}x^{l}}(0)=\frac{\textrm{d}^{l}u}{\textrm{d}x^{l}}(0),~~~~~~~~
\frac{\textrm{d}^{l}\widetilde{u}}{\textrm{d}x^{l}}(1)=\frac{\textrm{d}^{l}u}{\textrm{d}x^{l}}(1),~~~~~~~~
l=0,1,..,\frac{N}{2}-1.\label{eq-her-bound}
\end{eqnarray}
If the derivatives are known then $n= N-1$. When the derivatives are not known exactly on the grid they have to be approximated by finite difference methods, as done by Lalescu \textsl{et al.} \cite{Lalescu}. Unfortunately, this method is less accurate than Hermite interpolation and $n= N-2$.

The general framework will be illustrated with cubic Hermite interpolation for which $N=4$. So the interpolation uses the function value and the first derivative in the two neighboring grid points to construct the interpolation polynomial. We have chosen this method because it is very accurate. Moreover, the second derivative, which is a piecewise linear function, gives minimal errors with respect to the $L^2$-norm. This property is further discussed in Section \ref{sec_hermite}.

First, the discrete values of $u$ and possible derivatives which are given on the grid, are indicated with the vector $\textbf{b}$. In general we have
\begin{eqnarray}
\textbf{b}=\textbf{f}[u],\label{eq-sub0}
 \end{eqnarray}
 where the linear operator $\textbf{f}$ depends on the interpolation method and maps a function onto a $N$-vector.
Second, the coefficients $a_i$ of the monomial basis need to be computed. Because $\mathcal{I}$ and $\textbf{f}$ are linear operators, we can write without loss of generality,
\begin{eqnarray}
\textbf{a}^T=\textbf{b}^T\textbf{M}.\label{eq-sub1}
\end{eqnarray}
Here, $\textbf{M}$ is the matrix that defines the interpolation method. For illustration, $\textbf{f}$ and $\textbf{M}$ for cubic Hermite interpolation, are given by
\begin{eqnarray}
\textbf{f}[u]=\left( \begin{array}{c}
   u(0) \\
  \frac{\textrm{d}u}{\textrm{d}x}(0) \\
   u(1) \\
  \frac{\textrm{d}u}{\textrm{d}x}(1) \\
  \end{array}\right),~~~~~~~~\textbf{M}=\left( \begin{array}{cccc}
   1&0&-3&2 \\
   0&1&-2&1 \\
   0&0&3&-2 \\
   0&0&-1&1 \\
  \end{array}\right).
\end{eqnarray}
Finally, substituting relation ($\ref{eq-sub1}$) in ($\ref{eq-vector-a}$) gives
\begin{eqnarray}
\mathcal{I}[u](x)=\widetilde{u}(x)=\textbf{a}^T\bar{\textbf{x}}=\textbf{b}^T\textbf{M}\bar{\textbf{x}}.\label{eq-interpolation1d}
\end{eqnarray}
In many applications also derivatives are needed. In order to compute the $l$-th derivative of $\widetilde{u}$, the monomial basis functions should be differentiated $l$ times. In general this can be done by multiplying $\textbf{a}$ by the differentiation matrix $\textbf{D}$ $l$ times, so
\begin{eqnarray}
\textbf{a}^{(l)T}=\textbf{a}^T\textbf{D}^l,~~~~~~~~~~
\textbf{D} = \left( \begin{array}{ccccc}
  0&\cdots&\cdots&\cdots &0 \\
  1 & \ddots& &&\vdots \\
   0 &2  & \ddots&&\vdots\\
   \vdots &\ddots &  \ddots& \ddots&\vdots \\
  0 & \cdots &0 &N-1&0 \\
  \end{array}\right),
\end{eqnarray}
where $\textbf{a}^{(l)}$ contains the coefficients for the $l$-th derivative, obtaining
\begin{eqnarray}
\frac{\textrm{d}^l\widetilde{u}}{\textrm{d}x^l}(x)=\textbf{a}^{(l)T}\bar{\textbf{x}}=\textbf{b}^T\textbf{M}\textbf{D}^l\bar{\textbf{x}}=\textbf{b}^T\textbf{M}^{(l)}\bar{\textbf{x}},\label{eq-interpolation1d2}
\end{eqnarray}
with $\textbf{M}^{(l)}=\textbf{M}\textbf{D}^l$. Note that the matrix $\textbf{D}$ is nilpotent, since $\textbf{D}^l=\textbf{0}$ for $l\geq N$, implying that at most $N-1$ derivatives can be approximated.

In conclusion, we presented a framework that is able to describe interpolation methods, which can be used to implement the interpolation methods in a straightforward way. In Section \ref{sec_implementation} it is used to generate fast algorithms for the implementation of the method.

\section{3D interpolation}\label{sec_3d}

In this section the 1D interpolation methods of Section \ref{sec_met} are extended to the 3D case. Now the scalar field $u$ depends on the vector $\textbf{x}$ and a 3D interpolation needs to be carried out. Like before the interpolated field is given by $\widetilde{u}$ and $\mathcal{I}_3$ is the 3D equivalent of $\mathcal{I}$, so $\widetilde{u}=\mathcal{I}_3[u]$. The 3D field $u$ can be represented by a 3D Fourier series like
\begin{eqnarray}
u=\sum_\textbf{k}\hat{u}_\textbf{k}\phi_\textbf{k}(\textbf{x}),
\end{eqnarray}
where $\phi_\textbf{k}$ is given by
\begin{eqnarray}
\phi_\textbf{k}(\textbf{x})=e^{2\pi \textrm{i}\textbf{k}\cdot\textbf{x}}=\phi_{k_x}(x)\phi_{k_y}(y)\phi_{k_z}(z),
~~~~~~\textbf{k}=(k_x,k_y,k_z),~~\textbf{x}=(x,y,z),~~~~
\end{eqnarray}
and $\phi_k$ defined by (\ref{eq_phi_k}). Again we restrict ourselves to linear interpolation operators, therefore $\widetilde{u}$ can be written as
\begin{eqnarray}
\widetilde{u}(\textbf{x})=\sum_\textbf{k}\hat{u}_\textbf{k} \widetilde{\phi}_\textbf{k}(\textbf{x}),
\end{eqnarray}
with $\widetilde{\phi}_\textbf{k}$ the interpolant of $\phi_\textbf{k}$, i.e., $\widetilde{\phi}_\textbf{k}=\mathcal{I}_3\left[\phi_\textbf{k}\right]$.

\begin{figure}[!hbtp]
\centering
\includegraphics[width=8cm, viewport=.0in 7.5in 8.2in 11.5in]{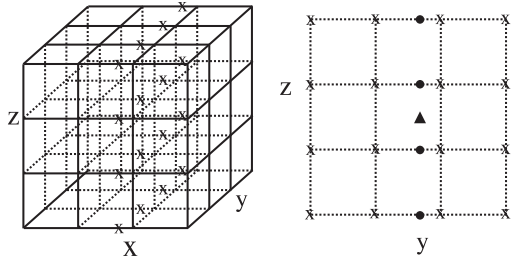}
\caption{Graphical description of the 3D Lagrange interpolation, using three steps of 1D interpolations for the case $N=4$. First, $N^2$ 1D interpolations are carried out in the $x$-direction (crosses). Second, $N$ interpolations are carried out in $y$-direction (dots in the right figure) and from these $N$ results finally one interpolated value is derived in $z$-direction (triangle). }\label{fig-3d-interpolatie}
\end{figure}

The 3D interpolation for a scalar field is carried out applying three times 1D interpolations, see Fig. \ref{fig-3d-interpolatie}. The interpolation consists of three steps, in which the three spatial directions are interpolated one after each other. The order in which the spatial directions are interpolated does not matter. Building the 3D interpolation out of 1D interpolations saves computing time. It can be done for all interpolation methods as long as the following two conditions are met. First, the operator $\mathcal{I}_3$ must be linear. Second, the following condition must be satisfied
\begin{eqnarray}
\widetilde{\phi}_\textbf{k}\equiv\mathcal{I}_3\left[\phi_\textbf{k}\right]=\mathcal{I}_3\left[\phi_{k_x}\phi_{k_y}\phi_{k_z}\right]=
\mathcal{I}\left[\phi_{k_x}\right]\mathcal{I}\left[\phi_{k_y}\right]\mathcal{I}\left[\phi_{k_z}\right]=\widetilde{\phi}_{k_x}\widetilde{\phi}_{k_y}\widetilde{\phi}_{k_z},
\label{eq-condition}
\end{eqnarray}
which is the case for almost all interpolation methods. Property (\ref{eq-condition}) can be used to prove that properties of the 1D operator $\mathcal{I}$ carry over to the 3D operator $\mathcal{I}_3$, for example, the order of convergence and the order of continuity.


Next, relations (\ref{eq-interpolation1d}) and (\ref{eq-interpolation1d2}) are extended to the 3D case. Like before, we start with storing some values of $u$ (given by the spectral code) and possible derivatives in the third-order tensor $\textbf{B}$. In the same fashion as relation (\ref{eq-sub0}) one gets
\begin{eqnarray}
\textbf{B}=\textbf{f}_z\Big[\textbf{f}_y\big[ \textbf{f}_x[u]\big]\Big],\label{eq-tensor-b}
\end{eqnarray}
where one element of tensor $\textbf{B}$ is defined like
\begin{eqnarray}
B_{i_1i_2i_3}=\textbf{f}_z\Big[\textbf{f}_y\big[ \textbf{f}_x[u]_{i_1}\big]_{i_2}\Big]_{i_3}.
\end{eqnarray}
$\textbf{f}_x$, $\textbf{f}_y$ and $\textbf{f}_z$ are similar to operator $\textbf{f}$ but now working in a specified direction. For Hermite interpolation they are given by
\begin{eqnarray}
\textbf{f}_x[u]=\!\!\left( \begin{array}{c}
   u (0,y,z)\\[0.1cm]
   \frac{\partial u}{\partial x} (0,y,z)\\[0.1cm]
   u (1,y,z)\\[0.1cm]
   \frac{\partial u}{\partial x} (1,y,z)\\
  \end{array}\right)\!,~~~~\textbf{f}_y[u]=\!\!\left( \begin{array}{c}
   u (x,0,z)\\[0.1cm]
   \frac{\partial u}{\partial y} (x,0,z)\\[0.1cm]
   u (x,1,z)\\[0.1cm]
   \frac{\partial u}{\partial y} (x,1,z)\\
  \end{array}\right)\!,~~~~\textbf{f}_z[u]=\!\!\left( \begin{array}{c}
   u (x,y,0)\\[0.1cm]
   \frac{\partial u}{\partial z} (x,y,0)\\[0.1cm]
   u (x,y,1)\\[0.1cm]
   \frac{\partial u}{\partial z} (x,y,1)\\
  \end{array}\right)\!.~~~~~~~~
\end{eqnarray}
The interpolation is carried out in a similar way as sketched in Fig. \ref{fig-3d-interpolatie}. Similarly to (\ref{eq-interpolation1d}), $\widetilde{u}(\textbf{x})$ can be represented as
\begin{eqnarray}
\mathcal{I}_3[u](\textbf{x})=\widetilde{u}(\textbf{x})=\textbf{B}\bar{\times}_1(\textbf{M}\bar{\textbf{x}})\bar{\times}_2(\textbf{M}\bar{\textbf{y}})\bar{\times}_3(\textbf{M}\bar{\textbf{z}}),\label{eq-interpolation}
\end{eqnarray}
where $\textbf{M}$ is still the matrix for 1D interpolation, $\bar{\textbf{y}}$ and $\bar{\textbf{z}}$ are defined like $\bar{\textbf{x}}$ which is given by relation (\ref{eq-vector-a}). Further, $\bar{\times}_n$ denotes the $n$-mode vector product \cite{tensor}, like
\begin{eqnarray}
\textbf{A}=\textbf{B}\bar{\times}_n\textbf{f},~~~~~~A_{i_1\cdots i_{n-1}i_{n+1}\cdots i_{\mathcal{N}}}=\sum_{i_n} B_{i_1\cdots i_{\mathcal{N}}} f_{i_n},
\end{eqnarray}
where $\mathcal{N}$ denotes the order of tensor $\textbf{B}$. In this way tensor $\textbf{A}$ is one order less than tensor $\textbf{B}$. Because we employ three of these  $n$-mode vector products the third-order tensor $\textbf{B}$ reduces to a scalar. Furthermore, each of these $n$-mode vector products corresponds to an interpolation in one direction, see also Fig. \ref{fig-3d-interpolatie}. For a general derivative one gets
 \begin{eqnarray}
\frac{\partial^{i+j+k}\widetilde{u}}{\partial x^i\partial y^j\partial z^k}(\textbf{x})= \textbf{B}\bar{\times}_1\left(\textbf{M}^{(i)}\bar{\textbf{x}}\right)\bar{\times}_2\left(\textbf{M}^{(j)}\bar{\textbf{y}}\right)\bar{\times}_3\left(\textbf{M}^{(k)}\bar{\textbf{z}}\right).\label{eq-interpolation2}
\end{eqnarray}
Note that the matrix $\textbf{M}$ does not necessarily have to be the same for the different directions $x$, $y$ and $z$. One could choose different interpolation methods when for example Chebyshev polynomials are used in one direction. In this case the grid is nonuniform in this direction and therefore not all interpolation methods can be used.

 Finally, when the scalar field $u(\textbf{x})$ becomes a vector field $\textbf{u}(\textbf{x})$, the three components of \textbf{u} can be interpolated separately. This can be written in short by a fourth order tensor $\textbf{B}$ where the last dimension contains the three components of $\textbf{u}$. In this way the equations for the new tensor $\textbf{B}$ remain the same as given above.
\section{Implementation}\label{sec_implementation}

 \begin{table}[!h]
 \caption{Algorithm for interpolation, with an estimate of the computational costs}
\centering
\begin{tabular}{ l c c}
\hline\hline
 Computed variables & Number of flops& Number of flops \\
 &&for $N\!=4$ \\
  \hline
  $\bar{\textbf{x}}$, $\bar{\textbf{y}}$ and $\bar{\textbf{z}}$&$3N$&12\\
  $\textbf{M}\bar{\textbf{x}}$, $\textbf{M}\bar{\textbf{y}}$ and $\textbf{M}\bar{\textbf{z}}$&$3N^2$&48\\
  $\textbf{M}^{(1)}\bar{\textbf{x}}$, $\textbf{M}^{(1)}\bar{\textbf{y}}$ and $\textbf{M}^{(1)}\bar{\textbf{z}}$&$3N(N-1)$&36\\
  $\textbf{B}\bar{\times}_1(\textbf{M}\bar{\textbf{x}})$&3$N^3$&192\\
  $\textbf{B}\bar{\times}_1\left(\textbf{M}^{(1)}\bar{\textbf{x}}\right)$&$3N^3$&192\\
  $\textbf{B}\bar{\times}_1(\textbf{M}\bar{\textbf{x}})\bar{\times}_2(\textbf{M}\bar{\textbf{y}})$&$3N^2$&48\\
  $\textbf{B}\bar{\times}_1(\textbf{M}\bar{\textbf{x}})\bar{\times}_2\left(\textbf{M}^{(1)}\bar{\textbf{y}}\right)$&$3N^2$&48\\
  $\textbf{B}\bar{\times}_1\left(\textbf{M}^{(1)}\bar{\textbf{x}}\right)\bar{\times}_2(\textbf{M}\bar{\textbf{y}})$&$3N^2$&48\\
  $\textbf{B}\bar{\times}_1(\textbf{M}\bar{\textbf{x}})\bar{\times}_2(\textbf{M}\bar{\textbf{y}})\bar{\times}_3(\textbf{M}\bar{\textbf{z}})$&$3N$&12\\
  $\textbf{B}\bar{\times}_1(\textbf{M}\bar{\textbf{x}})\bar{\times}_2(\textbf{M}\bar{\textbf{y}})\bar{\times}_3\left(\textbf{M}^{(1)}\bar{\textbf{z}}\right)$&$3N$&12\\
  $\textbf{B}\bar{\times}_1(\textbf{M}\bar{\textbf{x}})\bar{\times}_2\left(\textbf{M}^{(1)}\bar{\textbf{y}}\right)\bar{\times}_3(\textbf{M}\bar{\textbf{z}})$&$3N$&12\\
  $\textbf{B}\bar{\times}_1\left(\textbf{M}^{(1)}\bar{\textbf{x}}\right)\bar{\times}_2(\textbf{M}\bar{\textbf{y}})\bar{\times}_3(\textbf{M}\bar{\textbf{z}})$&$3N$&12\\
  \hline
  Total:&$6N^3+15N^2+12N$&672\\
  \hline\hline
\end{tabular}
\label{table-algoritm}
\end{table}

Relations (\ref{eq-interpolation}) and (\ref{eq-interpolation2}) provide a good starting point for an efficient implementation of the interpolation. We focus on interpolating a 3D vector field $\textbf{u}(\textbf{x})$ and on calculating all its first derivatives (which are needed in many applications like the computation of the trajectories of inertial particles). The matrices $\textbf{M}$ and $\textbf{M}^{(1)}$ only need to be computed once, which can be done prior to interpolation. Second, the vectors $\bar{\textbf{x}}$, $\bar{\textbf{y}}$ and $\bar{\textbf{z}}$ have to be computed which only needs to be done once for each position of interpolation. In Table \ref{table-algoritm} we keep track of all the computed quantities. Here, the computational costs for evaluating all the components is shown where one flop denotes one multiplication with one addition. We show the number of flops for the general case and for $N=4$. The main idea is to reduce the order of the tensors as soon as possible in order to generate an efficient method.

In order to determine how efficient the algorithm is, one can compare the computational costs against a lower bound. The lower bound we use is related to the size of $\textbf{B}$ which is $3N^3$ for a vector field $\textbf{u}$. In order to be able to use all the information in tensor $\textbf{B}$, $3N^3$ flops are needed. For large $N$ one finds that the algorithm of Table \ref{table-algoritm} is only a factor 2 less efficient than this lower bound.

We also compare our algorithm with the one proposed by Lekien and Marsden \cite{Lekien}, which uses Hermite interpolation with $N=4$. Our algorithm has less restrictions and shows a slightly better computational performance (for $N=4$). The algorithm of Lekien and Marsden consists of two steps. First, they calculate the coefficients for the polynomial basis. Second, the values at the desired location are calculated. They claim that their method is beneficial when the derivatives are needed or the interpolation needs to be done multiple times for the same interval, because only the second step needs to be executed multiple times. Our method does not have the first step, therefore it has no restrictions, nevertheless the computation of the values and the first derivatives is slightly faster than for Lekien and Marsden, even when considering only the second step. The total costs of their second step is bounded by $12N^3$ flops (4 times $3N^3$ flops, for the computation of the values and the first derivatives). From Table \ref{table-algoritm} we can conclude that our method needs less flops for the same computations.

\section{Fourier analysis}\label{sec_fourier}
In this section the interpolation operator $\mathcal{I}$ is expressed in terms of a convolution. In this way properties of the interpolation method like the order of continuity of the interpolated field and the magnitude of the errors can be shown in the Fourier domain. We start with the interpolation of 1D functions and subsequently, it can be extended to the 3D case.

\begin{figure}[!hbtp]
\centering
\includegraphics[width=13cm, viewport=.5in 2.9in 8.2in 11.5in]{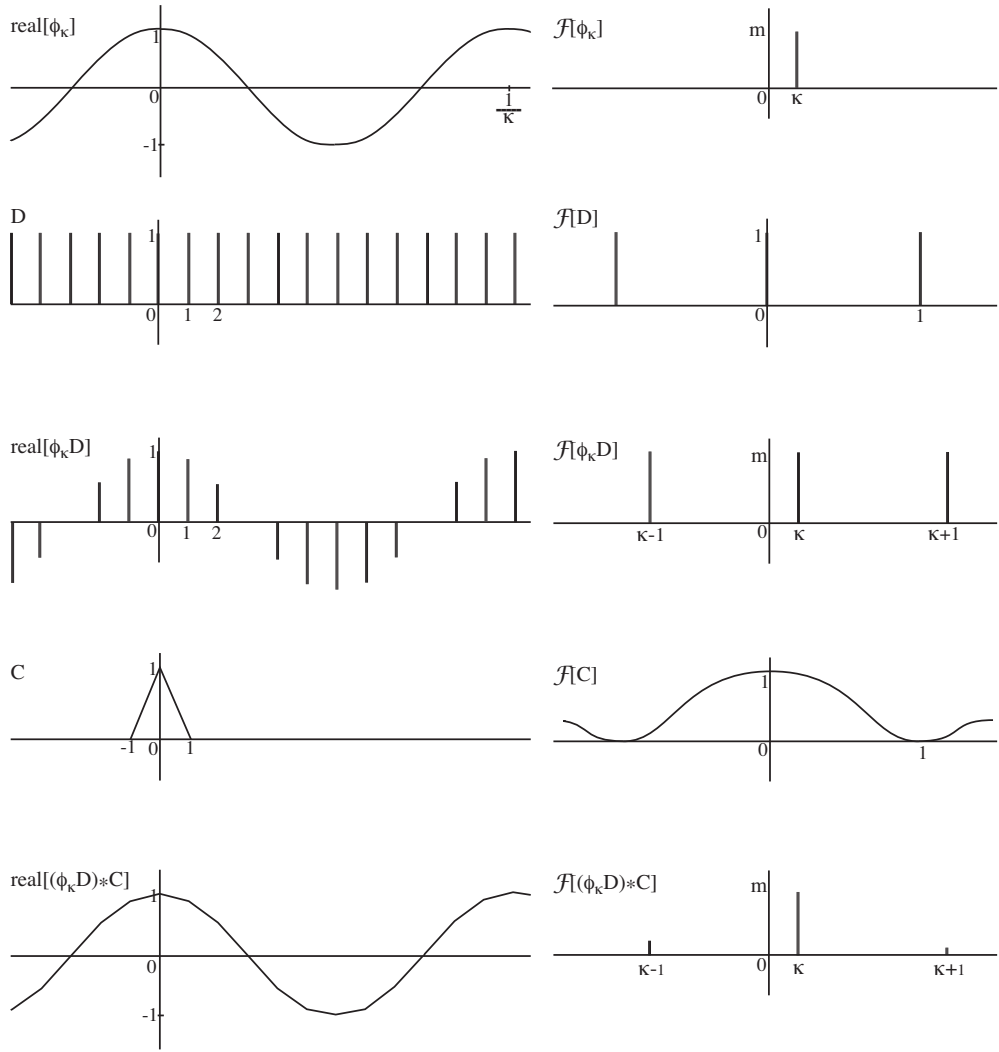}
\caption{Sketch of linear interpolation as a convolution. The pins represent delta functions with the height equal to its prefactor. On the left side is a visualization in real space and on the right side in Fourier space. }\label{fig-grafieken}
\end{figure}

Before we start with the derivation, we rescale the variable $x$ by dividing it by $\Delta x$, so that the new grid spacing equals unity. From now on we work with the rescaled grid where $x\in[0,m]$ and $m=\left(\Delta x\right)^{-1}$, so $x_j=j$ for $0\leq j\leq m$. Furthermore we introduce the dimensionless wave number $\kappa=k\Delta x$ and $\phi_{\kappa}$ is similarly defined as $\phi_k$, see (\ref{eq_phi_k}). For Hermite interpolation the derivation is somewhat more complex because also the derivatives are used and therefore it is postponed to Section \ref{sec_hermite}. We focus on interpolation methods where $\textbf{f}[u]$ contains the values of $u$ at the $N$ nearest grid points $x_j$ of $x$ with local ordering. Thus $b_j=u(x_j)$ and $x_j$ is given by
\begin{eqnarray}
x_j&=&\left\lfloor x-\frac{N}{2}+j\right\rfloor,~~~~~~~~~~~~j=1,2,\cdots,N,\label{eq-alg-int-6}
\end{eqnarray}
where $\lfloor\cdot \rfloor$ denotes the nearest lower integer. The interpolation methods can be described by the matrix $\textbf{M}$, with elements $M_{j,i}$, see relation (\ref{eq-interpolation1d}). This relation can also be written as
\begin{eqnarray}
\widetilde{u}(x)&=&\sum_{j=1}^NC_j\left(x-x_j\right)u\left(x_j\right),\label{eq-alg-int-3}
\end{eqnarray}
with $x_j$ defined by (\ref{eq-alg-int-6}) and where $C_j$ is given by
\begin{eqnarray}
C_j\left(x+\frac{N}{2}-j\right)&=&\left\{\begin{array}{cl}\sum_{i=1}^NM_{j,i}x^{i-1}&\textrm{for}~0\leq x<1,\smallskip\\
0&\textrm{elsewhere}.\end{array} \right.\label{eq-alg-int-1}
\end{eqnarray}
 Relation (\ref{eq-alg-int-3}) can be rewritten by using the sifting property of the delta function, like
\begin{eqnarray}
\widetilde{u}(x)&=&\sum_{j=1}^NC_j\left(x-x_j\right)\int_{-\infty}^{\infty}u(y)\delta\left(y-x_j\right)\textrm{d}y.\label{eq-alg-int-4}
\end{eqnarray}
This can be further reformulated by subtracting the argument of the delta function from the argument of $C_j$, as
\begin{eqnarray}
\widetilde{u}(x)&=&\int_{-\infty}^{\infty}u(y)\sum_{j=1}^N\delta\left(y-x_j\right)C_j(x-y)\textrm{d}y\nonumber\\
&=&\int_{-\infty}^{\infty}u(y)D(y)C(x-y)\textrm{d}y,\label{eq-alg-int-5}
\end{eqnarray}
with $C(x)$ and $D(x)$ given by
\begin{eqnarray}
C(x)=\sum_{j=1}^NC_j(x),~~~~~~~~~~
D(x)=\sum_{i\in\mathds{Z}}\delta(x-i).\label{eq-alg-int-2}
\end{eqnarray}
In relation (\ref{eq-alg-int-5}) the delta functions can be replaced by the function $D$, which is a train of delta functions because the functions $C_j$ only have a support of length unity, see (\ref{eq-alg-int-1}). Finally, the interpolation can be written like
\begin{eqnarray}
\widetilde{u}=(uD)*C,\label{eq-int}
\end{eqnarray}
with $*$ denoting the convolution product. Here, the convolution function $C$ depends on the interpolation matrix $\textbf{M}$, see Fig. \ref{fig-grafieken}.

As a consequence of relation (\ref{eq-int}), if the function $C$ is continuous up to the $p$-th derivative then $\widetilde{u}$ is also continuous up to the $p$-th derivative. Even stronger, the order of continuity of the function $C$ is equal to the order of continuity of $\widetilde{u}$. Furthermore, by the use of relation (\ref{eq-int}) exact interpolation can be constructed as well \footnote[1]{Exact interpolation can be accomplished by $\mathcal{F}[C](k)=1$ for $-0.5\leq k\leq0.5$ and zero elsewhere. In this way only the original Fourier component is filtered out of the spectrum. Note that in this case $C$ has infinite support.}.

In the following of this section we will discuss the interpolation error. Before proceeding we need to proof the following theorem.

\textbf{Theorem.} $\left\langle e_{\kappa},e_{\lambda}\right\rangle_2=0$ for $\kappa\neq\lambda$. Here $e_{\kappa}$ is the error in mode $\kappa$, $e_{\kappa}=\widetilde{\phi}_{\kappa}-\phi_{\kappa}$ and $\langle\cdot\rangle_2$ is the inner product related to the $L^2$-norm $\|\cdot\|_2$ defined by
\begin{eqnarray}
\langle f,g\rangle_2=\int_0^{m}\!\!f(x)g^*(x)\textrm{d}x,~~~~~~\left\|f\right\|_2^2=\langle f,f\rangle_2=\int_0^m\!\!f(x)f^*(x)\textrm{d}x.~~~~
\end{eqnarray}
The asterisk $(^*)$ denotes complex conjugation.

\textbf{Proof.} We start with replacing $u$ by $\phi_{\kappa}$ in  relation (\ref{eq-int}), i.e.,
\begin{eqnarray}
\widetilde{\phi}_{\kappa}=\mathcal{I}\left[\phi_{\kappa}\right]=\left(\phi_{\kappa}D\right)*C.\label{eq-fourier}
\end{eqnarray}
Second, we take the Fourier transform of $\widetilde{\phi}_{\kappa}$, for some fixed $\kappa_0$, i.e.,
\begin{eqnarray}
\mathcal{F}\left[\widetilde{\phi}_{\kappa_0}\right](k)&=&\mathcal{F}\Big[\left(\phi_{\kappa_0}D\right)*C\Big](k)=\Big(\mathcal{F}\left[\phi_{\kappa_0}\right]*\mathcal{F}[D]\Big)(k)\mathcal{F}[C](k)\nonumber\\
&=&m\sum_{i\in\mathds{Z}}\delta\big(k-(i+\kappa_0)\big)\mathcal{F}[C](i+\kappa_0),\label{eq-discrete}
\end{eqnarray}
which results in a train of delta functions with the perfector given by $\mathcal{F}[C]$, see Fig. \ref{fig-grafieken}, and $\mathcal{F}[\cdot]$ denotes the Fourier transform given by
\begin{eqnarray}
\mathcal{F}[g](k):=\int_{-\infty}^\infty g(x)e^{-2\pi \textrm{i}kx}\textrm{d}x.
\end{eqnarray}
For linear interpolation these functions are shown in Fig. \ref{fig-grafieken}. Next, $\langle e_{\kappa},e_{\lambda}\rangle_2$ can be written as $\langle e_{\kappa},e_{\lambda}\rangle_2=\left\langle \widetilde{\phi}_{\kappa},\widetilde{\phi}_{\lambda}\right\rangle_2-\left\langle \widetilde{\phi}_{\kappa},\phi_{\lambda}\right\rangle_2-\left\langle \phi_{\kappa},\widetilde{\phi}_{\lambda}\right\rangle_2+\left\langle \phi_{\kappa},\phi_{\lambda}\right\rangle_2$. Trivially $\left\langle \phi_{\kappa},\phi_{\lambda}\right\rangle_2=0$ for $\kappa\neq\lambda$. Furthermore, $\widetilde{\phi}_{\kappa}$ consists of a discrete set of Fourier components, see relation (\ref{eq-discrete}). Using this relation, one can show that no common Fourier components exist for $\widetilde{\phi}_{\kappa}$ and $\widetilde{\phi}_{\lambda}$ or $\phi_{\lambda}$ for $\kappa\neq\lambda$. Therefore $\left\langle \widetilde{\phi}_{\kappa},\widetilde{\phi}_{\lambda}\right\rangle_2=0$, $\left\langle \widetilde{\phi}_{\kappa},\phi_{\lambda}\right\rangle_2=0$ and $\left\langle\phi_{\kappa},\widetilde{\phi}_{\lambda}\right\rangle_2=0$ for $\kappa\neq\lambda$ implying $\langle e_{\kappa},e_{\lambda}\rangle_2=0$ as claimed.


\textbf{Corollary.} The orthogonality is important to estimate errors. When the error in $u$ is computed as $\|\widetilde{u}-u\|_2^2$, it can be rewritten like $\|\widetilde{u}-u\|_2^2=\sum_{\kappa}\hat{u}_{\kappa}^2\left\|e_{\kappa}\right\|_2^2$, which allows easy and straightforward computation of the errors.

Next, the error in one Fourier component is calculated. In this derivation we make use of the fact that $\widetilde{\phi}_{\kappa}$ can be written by a sum of Fourier components, see Fig. \ref{fig-grafieken} and relation (\ref{eq-discrete}). The relative error in one Fourier component can be written as
\begin{eqnarray}
\frac{\left\|\widetilde{\phi}_{\kappa}-\phi_{\kappa}\right\|_2^2}{\left\|\phi_{\kappa}\right\|_2^2}
=\frac{\left\|e_{\kappa}\right\|_2^2}{m}&=&\frac{1}{m}
\left\|-\phi_{\kappa}+\sum_{i\in\mathds{Z}}\mathcal{F}[C]\left(\kappa+i\right)\phi_{\kappa+i}\right\|_2^2\nonumber\\
&=&\big(\mathcal{F}[C]\left(\kappa\right)-1\big)^2+\sum_{i\neq0}\big(\mathcal{F}[C]\left(\kappa+i\right)\big)^2.\label{eq-l2-error}
\end{eqnarray}
From this expression one can see that the error can be computed directly from $\mathcal{F}[C]$. The same can be done for the error in the $l$-th derivative; $e^{(l)}_{\kappa}=\widetilde{\phi}^{(l)}_{\kappa}-\phi^{(l)}_{\kappa}$. The idea is to take the derivatives of the individual Fourier components which results in
\begin{eqnarray}
\frac{\left\|e^{(l)}_{\kappa}\right\|_2^2}{\left\|\phi^{(l)}_{\kappa}\right\|_2^2}=
\big(\mathcal{F}[C]\left(\kappa\right)-1\big)^2+\sum_{i\neq0}\left(\frac{\kappa+i }{\kappa}\right)^{2l}\big(\mathcal{F}[C]\left(\kappa+i\right)\big)^2.~~~~~~~~~~\label{eq-l2-error-der}
\end{eqnarray}

The extension to the 3D case is rather straightforward and is therefore not reported here. The basic idea is to create 3D functions by multiplying the 1D components, this can be done for all functions and the basic equations remain the same.
\section{Hermite interpolation}\label{sec_hermite}
In this section we extend the theory of Section \ref{sec_fourier} to Hermite interpolation. We also show some special properties that hold for Hermite interpolations. Especially, we examine the case $N=4$. For this case the second derivative becomes a piecewise linear function. Comparison with the actual second derivative shows that this piecewise linear function is optimal with respect to the $L^2$-norm.

In order to extend the theory of Section \ref{sec_fourier} to Hermite interpolation with even $N$ the same procedure is followed as in Section \ref{sec_fourier}. Analogous to (\ref{eq-alg-int-3}), $\widetilde{u}(x)$ can be written as
\begin{eqnarray}
\widetilde{u}(x)&=&\sum_{j=0}^1\sum_{l=1}^{N\!/2}C_{j,l}\left(x-x_j\right)\frac{\textrm{d}^{l-1}u}{\textrm{d}x^{l-1}}\left(x_j\right),
\end{eqnarray}
where $C_{j,l}$ and $x_j$ are given by
\begin{eqnarray}
C_{j,l}(x-j)&=&\left\{\begin{array}{cl}\sum_{i=1}^N
M_{l+j\frac{N}{2},i}x^{i-1}&\textrm{for}~0\leq x<1\\
0&\textrm{elsewhere}\end{array} \right.,~~~~~~~~l\in1,2,\cdots,\frac{N}{2},\nonumber\smallskip\\
x_j&=&\left\lfloor x\right\rfloor+j,~~~~~~~~~~\qquad\qquad\qquad\qquad~~~~~~~~~~~~~~~j\in0,1.
\end{eqnarray}
Again following similar steps as in Section \ref{sec_fourier}, $\widetilde{u}(x)$ can be rewritten as
\begin{eqnarray}
\widetilde{u}(x)&=&\sum_{j=0}^1\sum_{l=1}^{N\!/2}C_{j,l}(x-x_j)\int_{-\infty}^{\infty}\frac{\textrm{d}^{l-1}u}{\textrm{d}x^{l-1}}(y)\delta(y-x_j)\textrm{d}y\nonumber\\
&=&\sum_{l=1}^{N\!/2}\int_{-\infty}^{\infty}\frac{\textrm{d}^{l-1}u}{\textrm{d}x^{l-1}}(y)D(y)C_l(x-y)\textrm{d}y,
\end{eqnarray}
where $D$ is given by relation (\ref{eq-alg-int-2}) and $C_l$ is given by $C_l(x)=C_{0,l}(x)+C_{1,l}(x)$. In short, $\widetilde{u}$ can be written as
\begin{eqnarray}
\widetilde{u}=\sum_{l=1}^{N\!/2}\left(\frac{\textrm{d}^{l-1}u}{\textrm{d}x^{l-1}}D\right)*C_l,\label{eq-her-int-1}
\end{eqnarray}
similar to relation (\ref{eq-int}). Here one can see that for Hermite interpolation multiple convolution functions $C_l$ are needed which correspond to the derivatives and the function itself. Replacing $u$ by $\phi_{\kappa}$ in (\ref{eq-her-int-1}) gives
\begin{eqnarray}
\widetilde{\phi}_{\kappa}=\mathcal{I}
[\phi_{\kappa}]=(\phi_{\kappa}D)*\sum_{l=1}^{N/2}\left(2\pi\textrm{i}\kappa\right)^{l-1}C_l.
\end{eqnarray}
In this way we find a similar expression as relation (\ref{eq-fourier}), where $C$ has to be replaced by $\sum_{l=1}^{N/2}\left(2\pi\textrm{i}\kappa\right)^{l-1}C_l$. In conclusion, relation (\ref{eq-l2-error}) and (\ref{eq-l2-error-der}) can still be used.

\textbf{Property.}
For the error in the first derivative we have the following property:
\begin{eqnarray}
\left\langle e^{(1)},1\right\rangle_2=0,\label{eq-error-der}
\end{eqnarray}
where the inner product $\langle \cdot,\cdot\rangle_2$ is defined on the unit interval, i.e.,
\begin{eqnarray}
\left\langle f,g\right\rangle_2=\int_0^1f(x)g^*(x)\textrm{d}x.
\end{eqnarray}
Furthermore the error in the $l$-th derivative, $e^{(l)}$, is given by
\begin{eqnarray}
e^{(l)}=\frac{\textrm{d}^l\widetilde{u}}{\textrm{d}x^l}(x)-\frac{\textrm{d}^l u}{\textrm{d}x^l}(x).
\end{eqnarray}

\textbf{Proof.} One can rewrite, part of the interpolation conditions for Hermite interpolation (\ref{eq-her-bound}) in the following way
\begin{eqnarray}
\widetilde{u}(1)-\widetilde{u}(0)=u(1)-u(0)~~\Leftrightarrow~~\int_0^1\frac{\textrm{d}\widetilde{u}}{\textrm{d}x}\textrm{d}x=\int_0^1\frac{\textrm{d}u}{\textrm{d}x}\textrm{d}x.
\label{eq-conditions1}
\end{eqnarray}
Here two interpolation conditions give one new condition which is equivalent to relation (\ref{eq-error-der}).

\textbf{Corollary.} Property (\ref{eq-error-der}) shows that the error in the first derivative does not have a constant component. Therefore the constant component is exact with respect to the $L^2$-norm

\textbf{Property.} For the error in the second derivative in case of $N=4$ we have
\begin{eqnarray}
\left\langle e^{(2)},1\right\rangle_2=0,~~~~~~\left\langle e^{(2)},x\right\rangle_2=0.\label{eq-error-2der}
\end{eqnarray}

\textbf{Proof.} One can rewrite the interpolation conditions (\ref{eq-her-bound}) for $N=4$ in the following way
\begin{eqnarray}
\widetilde{u}(1)-\widetilde{u}(0)-\widetilde{u}'(0)=u(1)-u(0)-u'(0)&\Leftrightarrow&
\int_0^1\!\int_0^\alpha\frac{\textrm{d}^2\widetilde{u}}{\textrm{d}x^2}\textrm{d}x\textrm{d}\alpha=\int_0^1\!\int_0^\alpha\frac{\textrm{d}^2u}{\textrm{d}x^2}\textrm{d}x\textrm{d}\alpha,\nonumber\\
\widetilde{u}'(1)-\widetilde{u}'(0)=u'(1)-u'(0)&\Leftrightarrow&
\int_0^1\frac{\textrm{d}^2\widetilde{u}}{\textrm{d}x^2}\textrm{d}x=\int_0^1\frac{\textrm{d}^2u}{\textrm{d}x^2}\textrm{d}x.\label{eq-conditions}
\end{eqnarray}
The first relation in (\ref{eq-error-2der}) follows immediately from the second condition in (\ref{eq-conditions}). The second relation in (\ref{eq-error-2der}) is derived in the following way
\begin{eqnarray}
\int_0^1\!\int_0^\alpha\frac{\textrm{d}^2\widetilde{u}}{\textrm{d}x^2}\textrm{d}x\textrm{d}\alpha=\int_0^1\!\int_0^\alpha\frac{\textrm{d}^2u}{\textrm{d}x^2}\textrm{d}x\textrm{d}\alpha,~~~~~~~~\nonumber\\
\alpha\left.\int_0^\alpha\left(\frac{\textrm{d}^2\widetilde{u}}{\textrm{d}x^2}(x)-\frac{\textrm{d}^2u}{\textrm{d}x^2}(x)\right)\textrm{d}x\right|_{\alpha=0}^{\alpha=1}-\int_0^1\left(\frac{\textrm{d}^2\widetilde{u}}{\textrm{d}x^2}(\alpha)-\frac{\textrm{d}^2u}{\textrm{d}x^2}(\alpha)\right)\alpha\textrm{d}\alpha=0,~~~~~~~~\nonumber\\
\int_0^1\left(\frac{\textrm{d}^2\widetilde{u}}{\textrm{d}x^2}(x)-\frac{\textrm{d}^2u}{\textrm{d}x^2}(x)\right)x\textrm{d}x=0.~~~~~~~~
\end{eqnarray}
Here, the first step is integration by parts and the second step uses the second relation of equation (\ref{eq-conditions}).

\textbf{Corollary.} Relation (\ref{eq-conditions}) implies that $e^{(2)}$ does not have a constant component, neither a linear component. For $N=4$ the second derivative is a linear function, and this means that there is no better approximation in the $L^2$-norm of this second derivative with a piecewise linear function. This makes Hermite interpolation very interesting as a reference case, because we now have proven that the error is minimal for this case.

\section{B-spline interpolation}\label{sec_B_spline}
In this section we start with explaining B-spline interpolation. The idea is to create an as smooth as possible interpolant. Later it is shown how the pseudo-spectral code can be used to efficiently execute this interpolation method. Furthermore, the interpolation method is optimized to create small errors in the $L^2$-norm. We start with giving the B-spline convolution functions after which their matrix representation is given and finally the transformation to the B-spline basis functions is derived.

In a spectral code FFTs are applied to transform data from real space to Fourier space and backwards. These FFTs are the most expensive step in the simulation and therefore we want to keep the number of FFTs needed minimal. This is the reason why Hermite interpolation is not a good option, since extra FFTs are needed for the computation of the derivatives. An alternative is B-spline interpolation.

\begin{figure}[!hbtp]
\centering
\includegraphics[width=9cm]{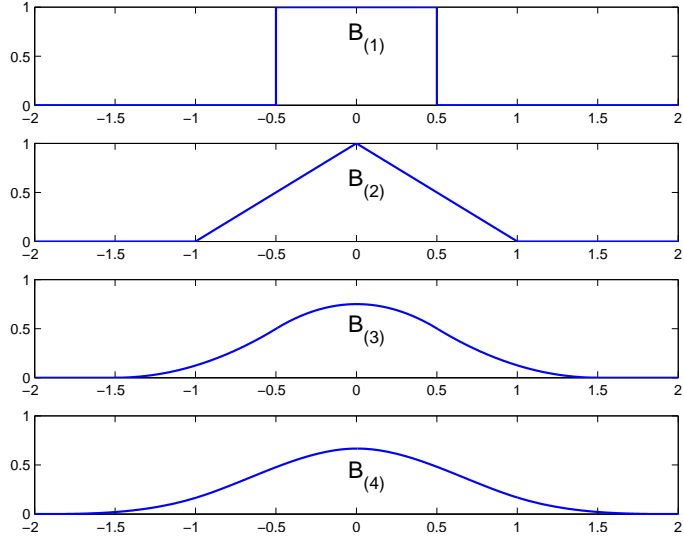}
\caption{First four uniform B-splines functions. }\label{fig-B_spline}
\end{figure}

We require high order of continuity of the interpolant. The highest order of continuity that can be obtained for the interpolant with piecewise polynomial functions of degree $N-1$ is $C^{N-2}$. In this way the interpolant still matches the original function $u(x)$ at the grid points $x_j$. Moreover, one can immediately see that  $n=N-1$, where $n$ is the highest degree of a polynomial test function for which the interpolation is still exact. This high level of continuity can be achieved by using B-spline functions \cite{Bspline2}. The first four uniform B-spline basis functions $B_{(N)}$ are shown in Fig. \ref{fig-B_spline}. These functions can be generated by means of convolutions in the following way
\begin{eqnarray}
B_{(1)}(x)&=&\left\{
 \begin{array}{ll}
 1~~~~&\textrm{for}~~~~-0.5\leq x<0.5,\\
0 &\textrm{elsewhere,} \\
  \end{array}
\right.\nonumber\\
B_{(2)}&=&B_{(1)}*B_{(1)},\nonumber\\
B_{(3)}&=&B_{(2)}*B_{(1)},\nonumber\\
&\vdots&\nonumber\\
B_{(N)}&=&B_{(N-1)}*B_{(1)}.
\end{eqnarray}
These functions have the property that the $N$-th function is of degree $N-1$ and is $C^{N-2}$. Furthermore, the B-spline basis functions have local support of length $N$. The B-spline functions can be seen as convolution functions $C$ introduced in Section \ref{sec_fourier} and have a matrix representation. The relation between the functions $B_{(N)}$ and the matrix representation is similar to relation (\ref{eq-alg-int-1}) and (\ref{eq-alg-int-2}), and is given by
\begin{eqnarray}
B_{(N)}(x)&=&\sum_{j=1}^{N}B_{(N),j}(x),\nonumber\\
B_{(N),j}\left(x+\frac{N}{2}-j\right)&=&\left\{\begin{array}{cl}\sum_{i=1}^{N}M_{(N),i,j}x^{i-1}&\textrm{for}~0\leq x<1,\\
0&\textrm{elsewhere.}\end{array} \right.
\end{eqnarray}
The matrix representation for the first four B-spline functions is as follows \cite{Bspline1}
\begin{eqnarray}
\textbf{M}_{(1)}&=&(1),\nonumber\\
\textbf{M}_{(2)}&=&\left(\begin{array}{cc}
 1&-1\\
0 &1 \\
\end{array}\right),\nonumber\\
\textbf{M}_{(3)}&=&\frac{1}{2!}\left(\begin{array}{ccc}
1&-2&1\\
1&2&-2 \\
0&0&1 \\
\end{array}\right),\nonumber\\
\textbf{M}_{(4)}&=&\frac{1}{3!}\left(\begin{array}{cccc}
1&-3&3&-1\\
4&0&-6&3 \\
1&3&3&-3\\
0&0&0&1 \\
\end{array}\right).
\end{eqnarray}
In general we have \cite{Bspline1}
\begin{eqnarray}
M_{(N),i,j}=\frac{1}{(N-1)!}Q_{N-1}^{N-j}\sum_{s=i}^{N}(-1)^{s-i}Q_{N}^{s-i}(N-s)^{N-j},~~~~~~~i,j=1,2,\cdots,N,~~~~~~\label{eq-B_spline_matrix}
\end{eqnarray}
with $Q_n^i$ given by
\begin{eqnarray}
Q_n^i=\frac{n!}{i!(n-i)!}=\left(\begin{array}{c}n\\i\end{array}\right).
\end{eqnarray}

We still need to express $u(x), x\in\mathds{Z}$ in terms of B-spline basis functions and thus find the transform from real space to the B-spline basis. Because the inverse transform from the B-spline basis to real space is somewhat easier, we start with this transformation first. From now on we omit the subindex $(N)$. The coefficients of the B-spline basis are called $u_B(x)$, and $u(x)$ can be derived from it by the discrete convolution $*_D$ in the following way, $u=u_B*_DB_D$.
Here, $B_D$ is given by
\begin{eqnarray}
B_D(x)=\left\{\begin{array}{ll}B(x)&\textrm{for}~x<\frac{m}{2}\\
B\left(x-m\right)&\textrm{for}~x\geq \frac{m}{2}\end{array} \right.,~~~~~~~~~~x=0,1,\cdots,m-1,
\end{eqnarray}
and the discrete convolution is given by
\begin{eqnarray}
\left(g*_Dh\right)(x)=\sum_{y=0}^{m} g(y)h\big((x-y)\!\!\!\!\mod m\big),~~~~~x=0,1,\cdots,m-1.~~~~~~
\end{eqnarray}
Next, the inverse $B_D^{-1}$ needs to be determined, where $B_D^{-1}$ is defined by
$B_D^{-1}*_DB_D=\delta_D$,
with $\delta_D$ the discrete delta function, given by
\begin{eqnarray}
\delta_D(x)=\left\{\begin{array}{cl}1&\textrm{for}~x=0\\
0&\textrm{else}\end{array} \right.~~~~~~~~~~x=0,1,\cdots,m-1.
\end{eqnarray}
Using the inverse $B_D^{-1}$, we can find $u_B(x)$ by the discrete convolution
$u_B(x)=u(x)*_DB_D^{-1}(x)$.

Using a spectral code, the discrete convolution can be evaluated in Fourier space and it reduces to a multiplication by constants $c(k)$. These multiplication constants can be computed beforehand and no convolutions need to be evaluated. In this way one gets
\begin{eqnarray}
\mathcal{F}_D\left[u_B\right](k)&=&\mathcal{F}_D[u](k)\mathcal{F}_D\left[B_D^{-1}\right](k)=
\frac{\mathcal{F}_D[u](k)}{\mathcal{F}_D\left[B_D\right](k)}\nonumber\\
&=&c(k)\mathcal{F}_D[u](k),\label{eq-B_spline_multiplication}
\end{eqnarray}
where the discrete Fourier transform $\mathcal{F}_D$ is given by
\begin{eqnarray}
\mathcal{F}_D[f](k)=\sum_{x=0}^{m-1}f(x)e^{-2\pi \textrm{i}xk/m}~~~~~~~k=0,1,\cdots,m-1.~~~~~~
\end{eqnarray}
The values of $c(k)$ can be determined in a straightforward manner as suggested above using $\mathcal{F}_D[B_D]$ but a more optimal choice for $c(k)$ can be made in the following way. We minimize the $L^2$-norm of the error and for this we use relation (\ref{eq-l2-error}) and we require
\begin{eqnarray}
\frac{\textrm{d}}{\textrm{d}c(k)}\left\|\phi_{\kappa}-c(k)\widetilde{\phi}_{\kappa}\right\|_2^2&=&0,
\end{eqnarray}
with $\kappa=k\Delta x$. This implies
\begin{eqnarray}
c(k)&=&\frac{\mathcal{F}[B]\left(\kappa\right)}{\sum_{i\in\mathds{Z}}\left(\mathcal{F}[B]\left(\kappa+i\right)\right)^2}.\label{eq-B_spline_coeff}
\end{eqnarray}
In three dimensions equation (\ref{eq-B_spline_multiplication}) becomes
\begin{eqnarray}
\mathcal{F}_D[u_B](\textbf{k})=c\left(k_x\right)c\left(k_y\right)c\left(k_z\right)
\mathcal{F}_D[u](\textbf{k}).\label{eq-B_spline_multiplication_3d}
\end{eqnarray}

Concluding, we propose an interpolation method for pseudo-spectral codes where the interpolation matrix $\textbf{M}$ is given by (\ref{eq-B_spline_matrix}). Further a multiplication in Fourier space is executed like (\ref{eq-B_spline_multiplication_3d}) where the coefficients can be determined from (\ref{eq-B_spline_coeff}). The coefficients can be computed beforehand and therefore no extra FFTs are needed, making this method very efficient.

\section{Comparison of the interpolation methods}\label{sec_comparison}
In this section four different interpolation methods are compared. The criteria we are interested in are the following. First, the method must be fast, which is needed because many interpolations will usually be carried out. Second, as we are using a spectral code, exponential convergence is expected and in order to meet this accuracy the interpolation methods must have high order of convergence. Furthermore, as the original function is $C^{\infty}$, the interpolated function must have a high order of continuity as well. Finally, the method must have small overall errors. In this way, also the derivatives of the interpolated field are still accurate enough.

The methods that are investigated are the following. First we have Lagrange interpolation where a polynomial function of degree $N-1$ passes through $N$ points \cite{Faires}. Second we have investigated the spline interpolation proposed by Lalescu \textsl{et al.} \cite{Lalescu}. Third, Hermite interpolation is considered and finally our newly proposed B-spline interpolation method is used. In Table \ref{table-comparison} some properties of the interpolation methods are reported. All the interpolation methods use piecewise polynomial functions of degree $N-1$ to reconstruct the field.

 \begin{table}[!h]
\centering
\begin{tabular}{ l c c c c}
\hline\hline
 method &$ n$ &order of&FFT&comment \\
 & &continuity&& \\
  \hline
 Lagrange interpolation&$N-1$ &$0  $&1&for even $N$\\
& &$-1  $&1&for odd $N$\\
 spline interpolation \cite{Lalescu}&$N-2$&$(N-2)/2$&1&only even $N$\\
 Hermite interpolation&$N-1$&$(N-2)/2$&$(N/2)^3$&only even $N$\\
 B-spline based interpolation&$N-1$&$N-2$&1&all $N$\\
  \hline
\end{tabular}
\caption{Overview of the interpolation methods investigated. For all methods the degree of the polynomial function is equal to $N-1$.}\label{table-comparison}
\end{table}

%

\begin{figure}[!hbtp]
\centering
\includegraphics[width=10cm]{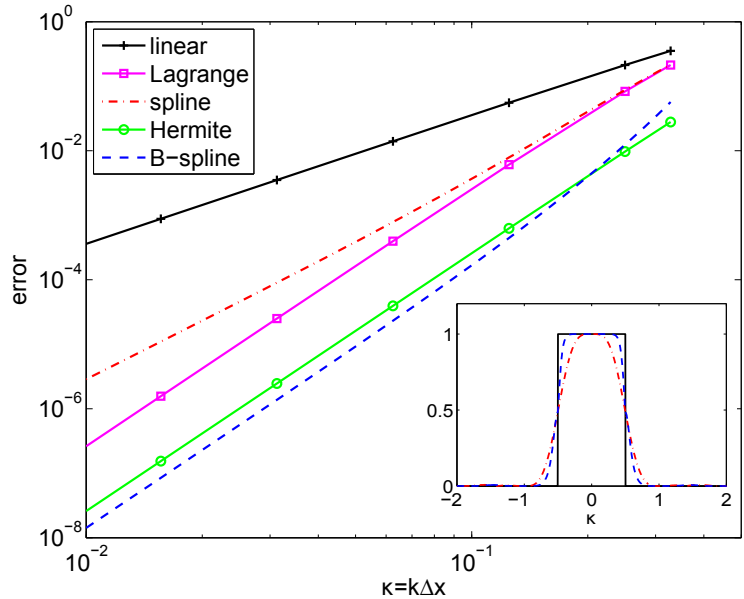}
\caption{Relative interpolation error for the Fourier mode, see Eq. (\ref{eq-l2-error}). For all methods $N=4$ except for linear interpolation which has $N=2$. The subfigure shows the Fourier transform of two interpolation kernels (spline and B-spline), where the solid line represents exact interpolation.}\label{fig-error-int}
\end{figure}

In order to estimate errors we use relation (\ref{eq-l2-error}) to find wave number dependent errors. For the four interpolation methods these errors are shown in Fig \ref{fig-error-int}. In our case $k_{max}\Delta x=\frac{1}{3}$ in order to avoid aliasing during the computation of the nonlinear term \cite{Spectral}. If this problem were not present, $k_{max}$ could be increased till $k_{max}\Delta x=\frac{1}{2}$. From Fig \ref{fig-error-int} also the order of convergence can be determined and it is found equal to $n+1$ (the lowest degree of a polynomial function for which the interpolation is not exact), in agreement with Table \ref{table-comparison}.

\begin{figure}[!hbtp]
\centering
\begin{minipage}[t]{0.48\linewidth}
\centering
\psfrag{kDx}{\scriptsize $\kappa=k\Delta x$}
\includegraphics[width=6.2cm]{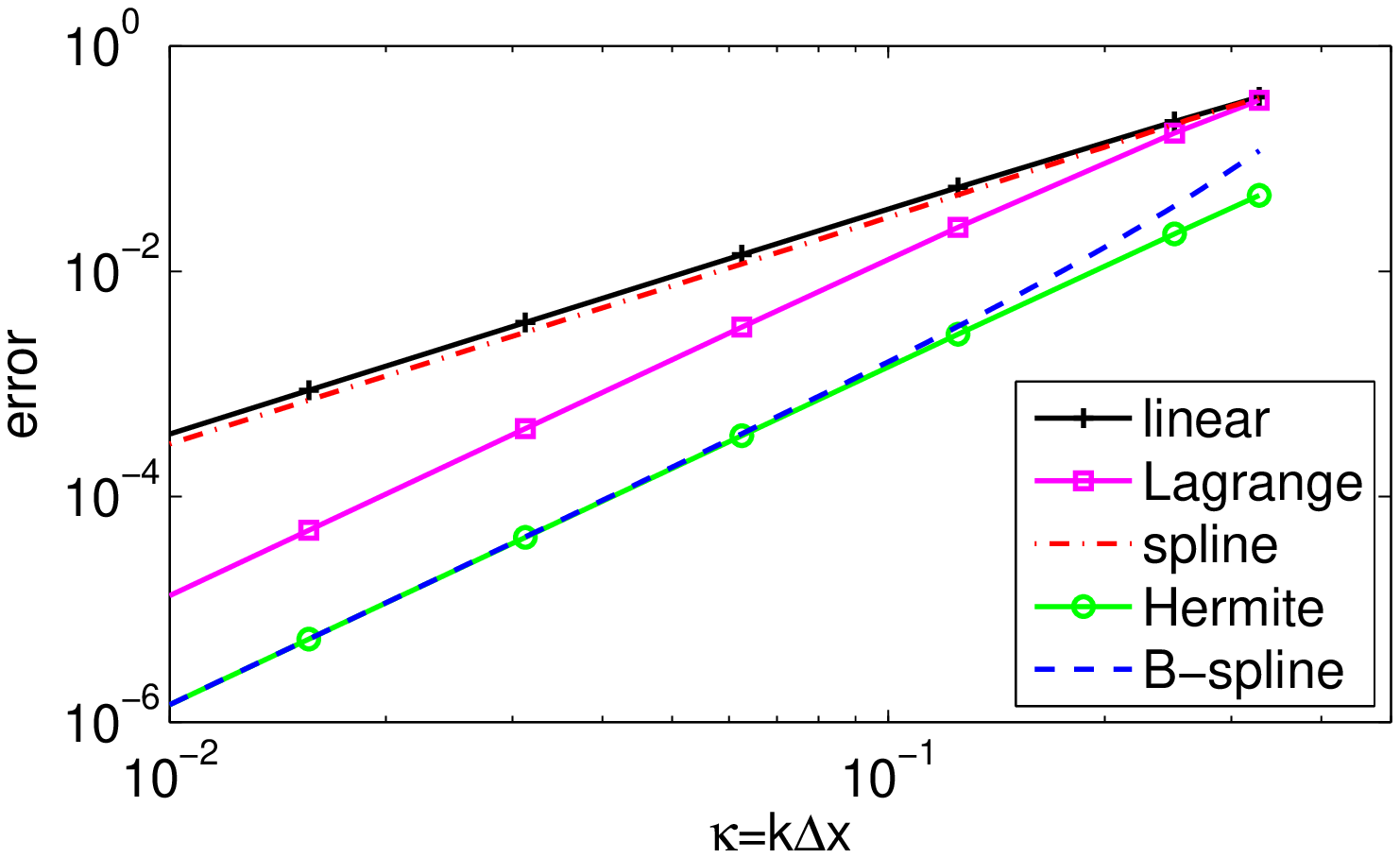}
\end{minipage}%
\hspace{0.5cm}%
\begin{minipage}[t]{0.48\linewidth}
\centering
\psfrag{kDx}{\scriptsize $\kappa=k\Delta x$}
\includegraphics[width=6.2cm]{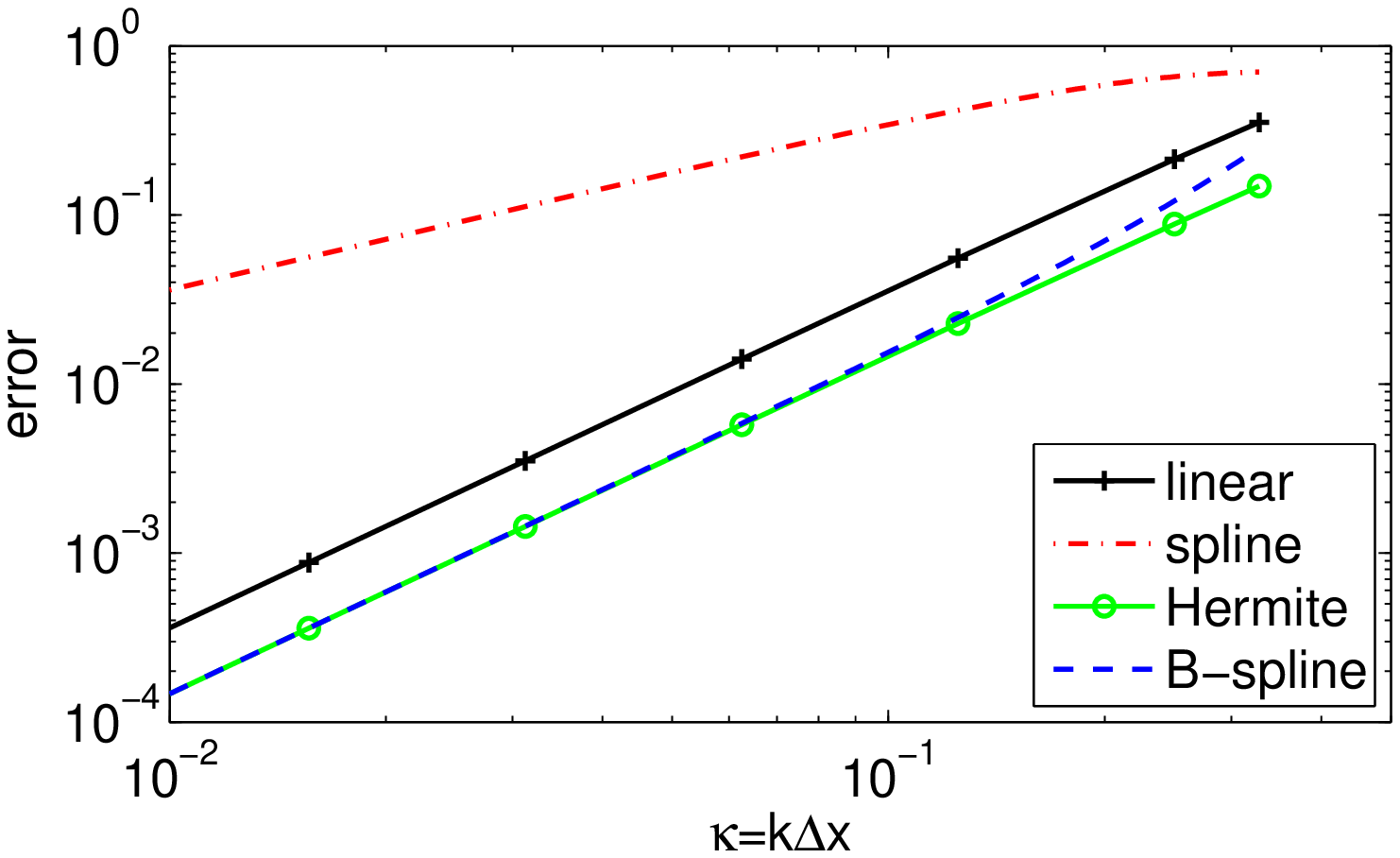}
\end{minipage}
\caption{Relative interpolation error for the first (left) and second derivative (right). Here the linear interpolation is taken on the first and second derivative where the other methods are taken on the function itself and then differentiated afterwards. Again all methods are taken with $N=4$ except for linear interpolation which has $N=2$.  }\label{fig-error-der}
\end{figure}

In order to avoid extra FFTs the interpolated field can be differentiated as done in relation (\ref{eq-interpolation2}) and the error can be computed by means of (\ref{eq-l2-error-der}). The interpolation errors of the first and second derivative are shown in Fig \ref{fig-error-der}. Here linear interpolation is executed on the derivatives themselves to give a comparison of how accurate the interpolation methods are. One can see for example that the second derivative is still better approximated by Hermite interpolation (with $N=4$) than with the linear interpolation executed on the second derivative.

When comparing the interpolation methods one can see that all interpolation methods have a weak point on one of our criteria except for the B-spline based method. The Lagrange interpolation for example is only $C^0$ continuous for even $N$ and even discontinuous for odd $N$. Furthermore, the overall error is relatively high compared with the other methods. The spline interpolation has already a better order of continuity but it has a lower order of convergence. Also the overall error is relatively high compared with the other methods. Hermite interpolation on the other hand has an excellent overall error, especially for the second derivative, see Section \ref{sec_hermite}. The main disadvantage of this method is that multiple FFTs are needed which is very time consuming. The B-spline based interpolation does not have this problem. The time it takes to execute the multiplication in Fourier space can be neglected compared with one FFT. Furthermore, this method reaches a much higher order of continuity compared with the other methods. When looking at the overall errors in Fig. \ref{fig-error-int} and \ref{fig-error-der} one can see that they almost match the one of Hermite interpolation. This is especially interesting for the second derivative because we have proven that there can not be a better approximation. Note that this second derivative is still continuous for the B-spline interpolation whereas for Hermite interpolation it is not.
\section{Conclusions}\label{sec_conclusion}
We have introduced a general framework for interpolation me-thods on a rectangular grid. Making use of this framework an algorithm is proposed for fast evaluation of the interpolation in three dimensions. This can easily save considerable computing time compared with other algorithms. It is shown that the computation time needed for this algorithm is close to a theoretical lower bound.

A spectral theory about these interpolation methods is presented, with which the spectral properties of the interpolation methods can be studied. Here basic properties of the interpolation method were shown like the order of continuity and the order of convergence. Furthermore, errors can be calculated for all Fourier components and also for its derivatives. By the use of this theory a novel B-spline based interpolation method is introduced for application in conjunction with spectral codes.

Finally, the interpolation methods for spectral codes are compared. The B-spline based interpolation method has several advantages compared with traditional me-thods. The order of continuity of the interpolated field is higher than that of Hermite interpolation and the other methods investigated. Second, only one FFT needs to be done where Hermite interpolation needs multiple FFTs for computing the derivatives. Third, the interpolation error matches almost the one of Hermite interpolations which is not reached by the other methods investigated. The proposed B-spline interpolation is thus the preferred candidate for particle tracking algorithms applied for turbulent flow simulations.

\section{acknowledgements}
We thank B.J.H. van de Wiel for fruitful discussions. The authors gratefully acknowledge financial support from the Dutch Foundation for Fundamental Research on Matter (FOM) (Program 112 "Droplets in Turbulent Flow"). This work was sponsored by the Stichting Nationale Computerfaciliteiten (National
Computing Facilities Foundation (NCF)) for the use of supercomputer facilities, with financial support from the Netherlands Organization for Scientific Research (NWO). The European COST Action MP0806 "Particles in Turbulence"  is also acknowledged.

\bibliography{bibliography}
       \bibliographystyle{unsrt}

\end{document}